\documentclass{article} 
\usepackage{nips13submit_e,times}
\usepackage{hyperref}
\usepackage{url}

\ifx\pdftexversion\undefined
\usepackage[dvips]{graphicx}
\else
\usepackage{graphicx}
\fi

\title{Interleaved Factorial Non-Homogeneous Hidden Markov Models for Energy Disaggregation}

\author{
Mingjun Zhong, Nigel Goddard, and Charles Sutton \\
School of Informatics \\
University of Edinburgh \\
Edinburgh, EH8 9AB \\
\texttt{\{mzhong,ngoddard,csutton\}@inf.ed.ac.uk}
}

\nipsfinalcopy 

\begin{document}

\maketitle

\begin{abstract}

  To reduce energy demand in households it is useful to know which
  electrical appliances are in use at what times.  Monitoring
  individual appliances is costly and intrusive, whereas data on
  overall household electricity use is more easily obtained.  In this
  paper, we consider the energy disaggregation problem where a
  household's electricity consumption is disaggregated into the
  component appliances. The factorial hidden Markov model (FHMM) is a
  natural model to fit this data. We enhance this generic model by introducing two
  constraints on the state sequence of the FHMM. The first is to use a non-homogeneous
  Markov chain, modelling how appliance usage varies over the day, and the other is to enforce that at most one chain changes state at each time step. 
 This yields a new model which we call 
   \emph{the interleaved factorial
  non-homogeneous hidden Markov model (IFNHMM)}. 
   We evaluated the
  ability of this model to perform disaggregation in an ultra-low
  frequency setting, over a data set of 251 English households.  
  In this new setting, the
  IFNHMM outperforms the FHMM in terms of recovering the
  energy used by the component appliances, due to that stronger
  constraints have been imposed on the states of the hidden Markov
  chains.  Interestingly, we find that the variability in model performance
  across households is significant, underscoring the importance of using
  larger scale data in the disaggregation problem.
\end{abstract}

\section{Introduction}

Two methods for reducing energy demand in households are to provide
occupants with detailed information about their energy use,
disaggregated into activities; and to use intelligent agents to
automate some aspects of energy use (e.g., when heating is turned on).
For both these purposes, it is important to identify which appliances are in use, and
when.  However, it is prohibitively expensive and
intrusive into householders' lives to monitor individual appliances on
a mass scale.  Methods to infer this data from the overall household
electricity signal are therefore of interest.  The literature,
however, does not yet provide a reliable method for performing
disaggregation in a large number of households.  Furthermore, previous
work uses high frequency data, readings one second apart or more
frequent, which can be difficult to obtain unobtrusively over long
periods.

In this paper we consider the energy disaggregation problem in an
ultra-low frequency setting (readings once every two minutes). We propose a new model,
called the interleaved factorial
non-homogeneous hidden Markov model (IFNHMM), which combines several
ideas from the literature, including the use of a non-homogenous HMM \cite{kim}
and a one-at-a-time constraint \cite{kolter}. We are unaware of previous
research which combines these two ideas into a single model.
We evaluate our model over a vary large and diverse dataset that to our knowledge
has not been used for this problem before, namely the UK Household Energy Survey
(HES) \cite{hes}, which studied 251 English households over the course of a year.
Interestingly, we find that the variation of the disaggregation error across
households is extremely high, of similar magnitude to the difference between our model and the baseline.
We suggest that addressing the variability of energy usage across
different households is a key challenge for future disaggregation research.

\emph{Problem Setting.} Suppose we have observed a time series of energy,
measured in watt hours by an electricity meter, denoted by
$Y=(Y_1,Y_2,\cdots,Y_T)$ where $Y_t\in{R_+}$. It is assumed that this
data were aggregated from the energy consumption of individual
appliances used by the household. Our aim is then to deconvolve the
aggregated data $Y$ to recover the energy which has been used by the
appliances individually. Suppose there were $I$ appliances and the
energy used by each appliance are denoted by
$X_i=(x_{i1},x_{i2},\cdots,x_{IT})$ where $x_{it}\in{R_+}$, so that we have the model:
\begin{equation}
Y_t = \sum_{i=1}^Ix_{it}
\label{eq1}
\end{equation}
The disaggregation problem is to recover the unknown time
series $X_{i}$ given the observed data $Y$.  This problem is essentially
a blind source separation (BSS) problem where only one mixture signal was observed
\cite{comon}, but this case is difficult for typical BSS algorithms.   The solution for the
model (\ref{eq1}) is not unique and due to the well-known
identifiability problem for the model (\ref{eq1}), i.e., 
 in that if the energy
used over the time step by any two appliances are equal, the two
appliances are exchangeable in the model.

\emph{Previous Work.} A review of different approaches to the disaggregation problem
can be found in \cite{zoha,zief}. A natural model for this problem, which has
been used in previous work on disaggregation, is the
factorial hidden Markov model (FHMM) \cite{Gha}.
Various approaches
have been proposed to improve the original FHMM for the disaggregation problem
\cite{kim,kolter,parson,Johnson}
Especially relevant to our work is Kolter and Jaakkolla \cite{kolter}, who use a one-at-a-time constraint for
the disaggregation problem, i.e., they allow most one hidden Markov chain to
change state at each time, and show that this assumption can be used to derive
specialized, efficient optimization algorithms.
Landwehr \cite{land} has used a version of the one-at-time-constraint
for activity recognition. Kim et al \cite{kim} take a slightly different approach
to allow the state transition probabilities to vary over time in the disaggregation
problem.  We follow the work on non-homogeneous hidden Markov chains by Diebolt et al
 \cite{diebolt}.

\section{Models}

The FHMM is a natural choice for modelling the generation of the total energy $Y$ where the energy used by each appliance is assumed to be a hidden Markov chain. Each hidden
 Markov chain is assumed to have a finite set of states, so that for each chain $i$, $x_{it}\in\{\mu_{i1},\cdots,\mu_{iK_i}\}$ where $K_i$ denotes the number of the states of chain $i$. In the FHMM, the model parameters which are denoted by $\theta$ are actually unknown, and those parameters are the initial probabilities $\pi_i=(\pi_{i1},\cdots,\pi_{iK_i})^T$ for each chain, the transition probabilities $p^{(i)}_{jk}=P(S_{it}=j|S_{i,t-1}=k)$ where $S_{it}\in\{1,2,\cdots,K_i\}$ denotes the states for the chain $i$, and the $\mu_{ik}$. Those parameters could be estimated by using approximation methods such as the structural variational approximation \cite{Gha}. However, in this paper we focus on inferring the hidden states $S_{it}$ for each hidden Markov chain, and simply estimate the parameters via maximum likelihood on supervised 
 training data in which the disaggregated energy signals $x_{it}$ have been measured.
 
We extend the basic FHMM in two ways.
First, we believe that some appliances are rarely used in some periods of time, depending on the behaviour of the household occupants. Therefore, each household has its own pattern for using the appliances. For example the dishwasher may never be used in the morning, and thus is always in the OFF state during that period. In that case the transition probability to the ON state will be zero during the morning. Given that the basic FHMM assumes that the hidden Markov model is homogeneous where the transition probabilities are always the same at any time, it is natural to assume that the transition probabilities from time $t$ to $t-1$ are changing. So we assume a factorial non-homogeneous hidden Markov model and denote the transition probabilities at time $t$ by $p^{(i)}_{tjk}$.

Second, we impose a further constraint on the states $S$, namely, that from time $t$ to $t+1$, we expect that a small number of hidden chains can change state. In this paper, we study the case that at most one chain is allowed to change state from the current time to the next. We call this model the interleaved factorial hidden Markov model (IFHMM), following Landwehr \cite{land}. Furthermore, under this constraint we could also assume the non-homogeneous Markov chain, which is thus the interleaved factorial non-homogeneous hidden Markov model (IFNHMM). For the IFHMM and IFNHMM an extra hidden Markov chain $Z\in\{1,2,\cdots,I\}$ is added to indicate which chain is permitted
to change state. All the other chains are required to remain in their current state. Generally, the IFNHMM has the following joint distribution\begin{equation}
p(Y,S,Z)=P(Z_1)\prod_{i=1}^IP(S_{i1})\prod_{t=1}^TP(Z_t|Z_{t-1})p(Y_t|S_t)\prod_{i=1}^IP_t(S_{it}|S_{i,t-1},Z_t)
\end{equation}
where $p(Y_t|S_t)=N(\sum_{i=1}^T\mu_{i,S_{it}},\sigma^2_t)$ is a Gaussian distribution, and $P_t$ is the time varying transition probability where $P_t(S_{it}=j|S_{i,t-1}=k,Z_t=i)=p^{(i)}_{tjk}$ and $P_t(S_{it}=j|S_{i,t-1}=k,Z_t\neq i)=\delta\{j=k\}$. Note that the IFHMM has the similar joint distribution and the only difference is that the Markov chains are homogeneous. To infer $X_i$, it is required to infer the optimal hidden states $\{S^*, Z^*\}$ such that
\begin{equation}
\{S^*, Z^*\}=\arg\max_{S,Z}P(S,Z|Y).
\end{equation}
For the FHMM and FNHMM the basic chainwise Viterbi algorithm is used to optimize those hidden states. For the IFHMM and IFNHMM, to infer the optimal hidden states we need to optimize the posterior with respect to any of the two chains $S_m$ and $S_n$, such that
\begin{equation}
\{S_m^*,S_n^*, Z^*\}=\arg\max_{S_m,S_n,Z}P(S,Z|Y).
\end{equation}
A chainwise Viterbi algorithm can be derived, similarly to Landwehr \cite{land}.

\section{Data and Results}
We apply our models to the Household
Electricity Survey (HES) data, a recent study commissioned by the UK Department of Food and Rural Affairs, which monitored a total of 251
owner-occupier households across the England from May 2010 to July
2011 \cite{hes}.  The study monitored 26 households for an entire year, while the remaining
225 were monitored for one month during the year with periods selected
to be representative of the different seasons.  Individual appliances
as well as the overall electricity consumption were monitored.  The
households were carefully selected to be representative of the overall
population.  The data
were recorded every 2 or 10 minutes, depending on the household.
This \emph{ultra-low frequency} data presents a challenge for disaggregation techniques;
typically studies rely on much higher data rates, e.g., the REDD data
\cite{kolter}. In our evaluation we employed data from 100 of the households, which were recorded in
every 2 minutes, to compare the four models. To estimate the model
parameters, 20 to 30 days data were used and another 5 to 10 days data
which were not using as the training data were used to compare those
models.

After the states were inferred for each chain, we measure the disaggregation performance using the normalized squared error 
$E = {\sum_{i,t}(\hat{x}_{it}-x_{it})^2}/{\sum_{i,t}x_{it}^2},$ where $\hat{x}_{it}$ denotes the estimated energy in power used by the appliance $i$ at time $t$.  
The results for each model applying to one hundred households are shown in Table 1. 
We observe that the more complex models perform better, and our final model, the IFNHMM, performed best.
This suggests that the identifiability problem was slightly reduced given the constraints imposed on the hidden state sequences.  It is especially interesting
to examine the parenthesized numbers in Table 1, which indicate the standard
deviation of the disaggregation errors across households.  The standard deviations are high---the coefficient of variation for the IFNHMM error is 0.389---and are of similar
magnitude to the difference in error between the different models.
This suggests that the \emph{variability} in energy usage across households
is a key research issue for home energy disaggregation, and one that
cannot be explored on the smaller scale data sets that have been most
commonly used in disaggregation studies to date.

As an example, in Figure 1 we plot, for a few appliances, the true power and the power  inferred by using the IFNHMM. We show results on
two different households. Obviously, the method could perform well in some houses for example the left two columns, 
but perform badly in some other houses for example the right two columns due to the identifiability problems. 


\section{Conclusions}
We have applied four factorial hidden Markov models to one hundred household data 
to recover the energy used by every appliance. Two constraints were made on the hidden Markov chains,
 where one is the non-homogeneous assumption on the chain and the other is to assume that 
at most one chain to change its state at any time. By comparing the total disaggregation error 
computed from the four models, the IFNHMM performed best among the other methods. 
This shows that incorporating additional constraint into the model helps to reduce the identifiability problem. 

\section*{Acknowledgements}

This work was supported by the Engineering and Physical Sciences
Research Council (grant number EP/K002732/1).

\begin{figure}[h]
\begin{center}
{\includegraphics[width=0.8\linewidth]{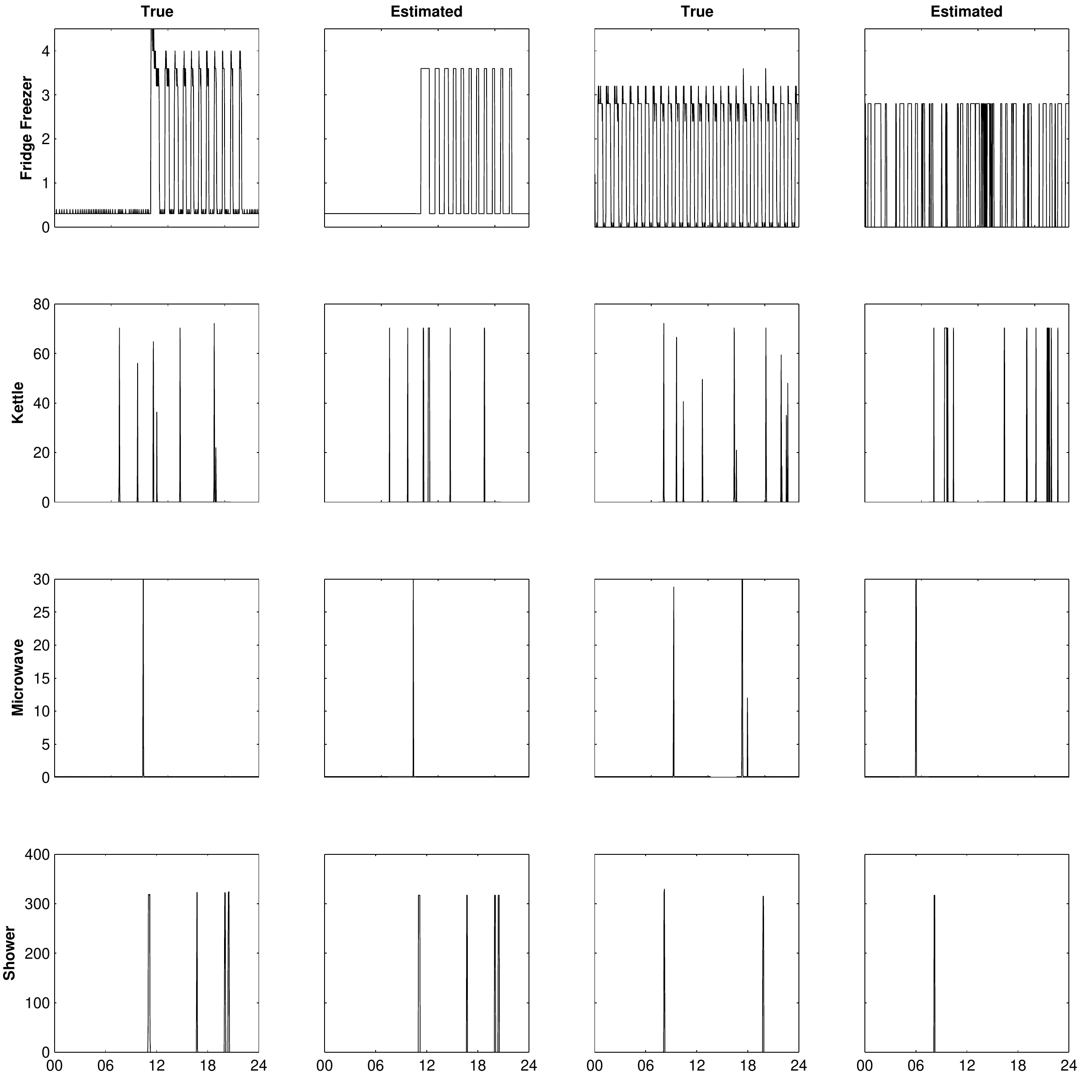}}
\end{center}
\caption{The true energy over one day for each appliance and the disaggregated power by the IFNHMM method, for household A (left two columns) and household B (right two columns).  Horizontal axes are in hours, vertical axes in watt hours, and measurements are accumulated over every two minutes.}
\end{figure}


\begin{table}[t]
\caption{Total Reconstruction Error computed by FHMM, FNHMM, IFHMM and IFNHMM for one hundred households. The values are shown in mean(std).}
\label{sample-table}
\begin{center}
\begin{tabular}{llll}
\multicolumn{1}{c}{\bf FHMM} &\multicolumn{1}{c}{\bf FNHMM} & \multicolumn{1}{c}{\bf IFHMM} &\multicolumn{1}{c}{\bf IFNHMM}
\\ \hline 
1.024 (0.396) & 0.947 (0.368) & 0.892 (0.334) & 0.847 (0.289)\\

\end{tabular}
\end{center}
\end{table}

\newpage

\small{

}

\newpage


\begin{thebibliography}{1}

  \bibitem{comon} Comon, P. \& Jutten, C. (2010). {\it Handbook of Blind Source Separation: Independent Component Analysis and Applications.} Academic Press.
  
  \bibitem{diebolt}Diebolt, F.X., Lee, J.H., \& Weinbach, G.C. (1994). Regime switching with time varying transition probabilities. In: C.P.Hargreaves (Ed.) {\it Nonstationary Time Series Analysis and Cointegration} (Oxford: Oxford University Press), pp. 283-302.  
  
  \bibitem{Gha} Ghahramani, Z.\& Jordan, M.I. (1997). Factorial hidden Markov models. {\it Machine Learning}, 27:245-273.

  \bibitem{Hart}Hart, G. (1992). Nonintrusive appliance load monitoring. {\it Proceedings of the IEEE}, 80(12)

  \bibitem{kim}Kim, H., Marwah, M., Arlitt, M., Lyon, G. \& Han, J. (2011). Unsupervised disaggregation of low frequency power measurements. {\it Proceedings of the SIAM Conference on Data Mining.}

  \bibitem{Johnson}Johnson, M.J., \& Willsky, A.S. (2013). Bayesian nonparametric hidden semi-Markov models. {\it Journal of Machine Learning Research}, 14:673-701.

  \bibitem{kolter10}Kolter, J.Z., Batra, S. \& Ng,A.Y. (2010) Energy disaggregation via discriminative sparse coding. {\it Neural Information Processing Systems.}

  \bibitem{kolter} Kolter, J.Z., \& Jaakkola, T. (2012) Approximate inference in additive factorial HMMs with application to energy disaggregation. {\it Appearing in Proceedings of the 15th International Conference on Artificial Intelligence and Statistics.}

  \bibitem{land}Landwehr, N. (2008). Modeling interleaved hidden processes. {\it Proceedings of the 25th International Conference on Machine Learning.}

  \bibitem{parson}Parson, O., Ghosh, S., Weal, M., \& Rogers, A. (2012) Non-intrusive load monitoring using prior models of general appliance types. AAAI, 2012.   
   
  \bibitem{saul}Saul, L.K. \& Jordan, M.I. (1999). Mixed memory Markov chains: decomposing complex stochastic processes as mixtures of simpler ones. {\it Machine Learning}, 37: 75-87.  
     
  \bibitem{zief} Ziefman, M.\& Roth, K.(2011). Nonintrusive appliance load monitoring: review and outlook. {\it IEEE transactions on Consumer Electronics}, 57(1): 76-84.  
  
  \bibitem{zoha} Zoha, A., Gluhak, A., Imran, M.A., \& Rajasegarar, S. (2012). Non-intrusive load monitoring approaches for disaggregated energy sensing: a survey. {\it Sensors}, 12: 16838-16866.
  
  \bibitem{hes} Zimmermann, J.-P., Evans, M., Griggs, J., 
King, N., Harding, L., Roberts, P., and Evans, C.   
Household Electricity Survey. Department of Food and Rural Affairs UK.
2012. \url{http://randd.defra.gov.uk/Default.aspx?Menu=Menu&Module=More&Location=None&Completed=0&ProjectID=17359}                             
    
\end{thebibliography}
\end{document}